\documentclass[%
 reprint,
superscriptaddress,
 amsmath,amssymb,
 aps,
prl,
]{revtex4-1}

\usepackage{graphicx}
\usepackage{dcolumn}
\usepackage{bm}
\usepackage{bbold}
\usepackage{float}

\begin{document}

\preprint{APS/123-QED}

\title{Model-based machine learning of critical brain dynamics}


 \author{Hernán Bocaccio}%
\affiliation{Physics Department, University of Buenos Aires, and Buenos Aires Physics Institute, Argentina}%
 \author{Enzo Tagliazucchi}%
\affiliation{Physics Department, University of Buenos Aires, and Buenos Aires Physics Institute, Argentina}%
\affiliation{Latin American Brain Health Institute (BrainLat), Universidad Adolfo Ibañez, Santiago, Chile}%

\date{\today}

\begin{abstract}
Criticality can be exactly demonstrated in certain models of brain activity, yet it remains challenging to identify in empirical data. We trained a fully connected deep neural network to learn the phases of an excitable model unfolding on the anatomical connectome of human brain. This network was then applied to brain-wide fMRI data acquired during the descent from wakefulness to deep sleep. We report high correlation between the predicted proximity to the critical point and the exponents of cluster size distributions, indicative of subcritical dynamics. This result demonstrates that conceptual models can be leveraged to identify the dynamical regime of real neural systems.

\end{abstract}

\maketitle

The criticality hypothesis states that neural activity is posed at or near the critical point of a second order phase transition \cite{chialvo2010emergent, cocchi2017criticality}. Theoretical support for this hypothesis is found in the information processing benefits that result from operating near the critical point, such as increased dynamic range, capacity for information transmission, and information capacity \cite{kinouchi2006optimal, shew2013functional, beggs2008criticality}. At the critical point, the repertoire of possible states is maximized, together with the susceptibility and correlation length \cite{haimovici2013brain, expert2011self, hahn2017spontaneous}. In combination, these properties result in highly differentiated dynamics, yet also integrated and reactive, as considered necessary to support higher cognitive function and conscious awareness \cite{tononi2004information, mashour2020conscious}. 

Although criticality presents distinct signatures, establishing it from neural activity recordings remains challenging \cite{beggs2012being}. Experiments using an ample variety of recording techniques have found evidence of scale-free neural activity avalanches, a landmark feature of self-organized criticality \cite{bak1988self, beggs2003neuronal, priesemann2014spike, tagliazucchi2012criticality, shriki2013neuronal, scott2014voltage}. However, it is difficult to demonstrate power law relationships with a limited amount of noisy data; moreover, there are other mechanisms consistent with scale-free behavior besides criticality \cite{girardi2021brain}. Other possible signatures of criticality are based on branching ratios \cite{poil2008avalanche}, spatio-temporal correlations \cite{expert2011self, haimovici2013brain}, and the response to external perturbations \cite{tagliazucchi2016large}. While these alternative metrics have been explored with success, establishing a single unified criterion for the presence of criticality in neural systems remains an open problem. 

In contrast to experimental data, criticality can be rigorously demonstrated in mathematical models of physical systems, including some models used to represent neural activity, such as those based on self-organized criticality \cite{bak1988self}. Even in models lacking neurobiological realism, the universality of dynamics near criticality could ensure the statistical similarity between real and simulated activity  \cite{chialvo2010emergent, cocchi2017criticality}. We explored this idea by implementing a model known to exhibit a second order phase transition (Greenberg-Hastings cellular automata) \cite{haimovici2013brain} coupled with the anatomical connectome, obtained using diffusion spectrum imaging (DSI) \cite{hagmann2008mapping}. We trained a deep neural network to output the probability of the model operating at the sub- and supercritical regimes, which allowed us to correctly identify the critical point as the crossing point between both probabilities. Next, we showed that the probabilities obtained from the neural network applied to empirical functional magnetic resonance imaging (fMRI) data presented high correlation with the power-law exponents of avalanche distributions obtained from the same data across different brain states (wakefulness and all stages of the human non-rapid eye movement [NREM] sleep cycle). Thus, we showed that the proximity to critical dynamics can be estimated from multivariate experimental data by means of a machine learning algorithm trained using simulations of a conceptual model of brain activity exhibiting criticality. 

\begin{figure*}
\includegraphics[scale=0.26]{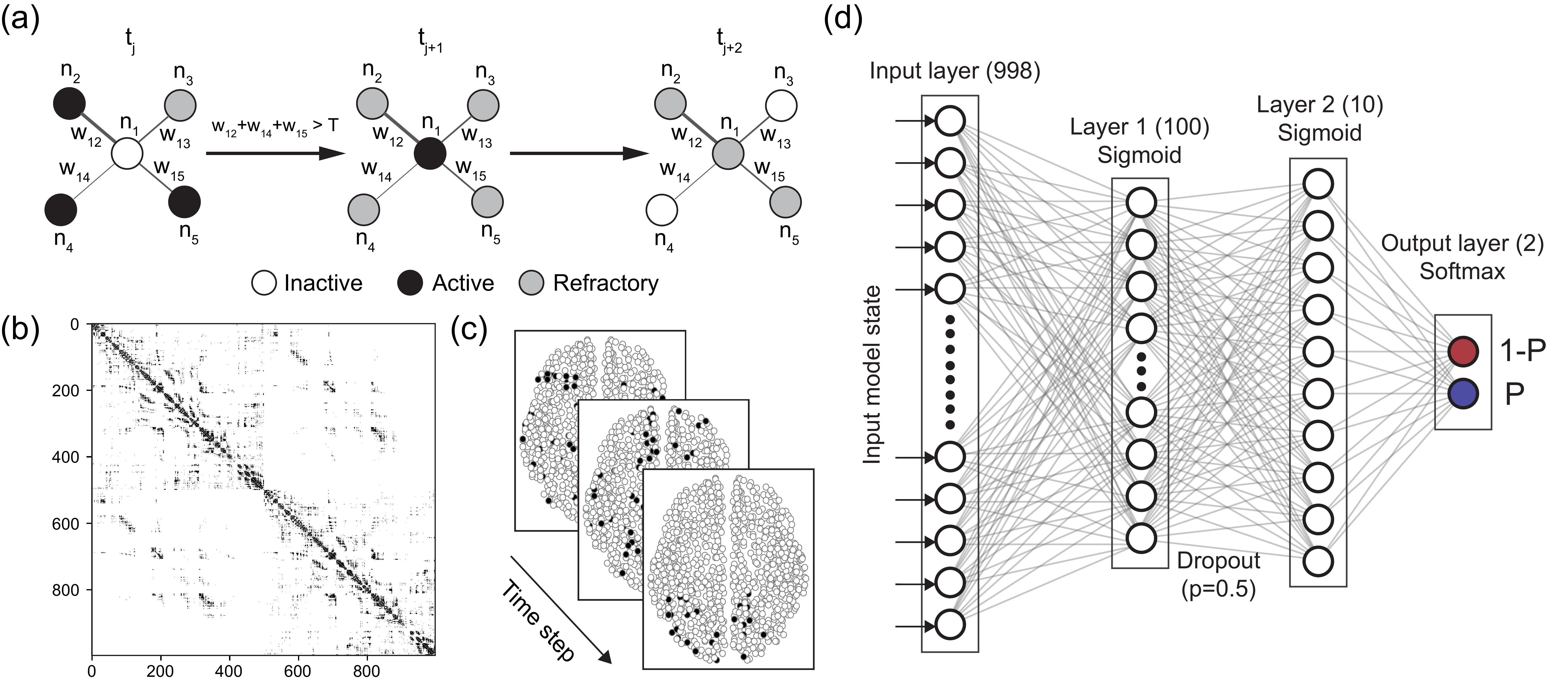}
\caption{\label{method} Methodological overview. (a) Transition rules for the local dynamics in the Greenberg-Hastings model, where $t_i$ is the $i$-th discrete time step, $n_i$ is the $i$-th node, $W_{ij}$ is the connection weight between nodes $i,j$, and $T$ is the threshold for the propagation of activity. (b) Matrix representation of the anatomical connectome used in the model, corresponding to 998 nodes densely connected by weights estimated using DSI. (c) Example of the model output evolving over time. (d) Diagram of the neural network architecture used to learn the proximity to criticality in the model, consisting of one input layer (998 units, one per node in the model), followed by two internal layers (100 and 10 units, respectively) with sigmoid activation functions, densely connected with 0.5 dropout probability, and by an output layer with two softmax units.}
\end{figure*}

\begin{figure}
\includegraphics[scale=0.16]{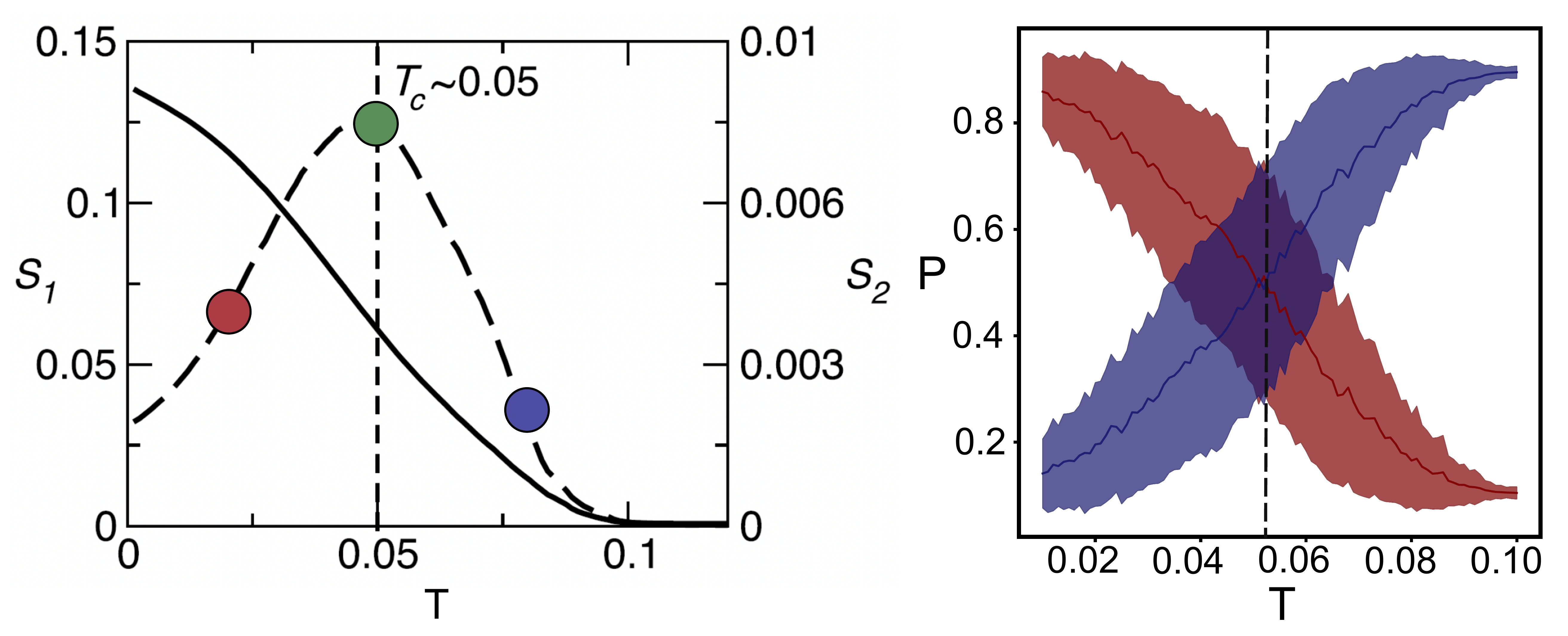}
\caption{\label{phase} Learning the phase transition. Left: Phase transition in the model (adapted from \cite{haimovici2013brain}). Here $S_1$ and $S_2$ stand for the largest and second largest connected clusters of active nodes. The red, blue and green dots indicate sub-, super-, and critical points of the model, with the critical threshold $T_c \approx 0.05$. Right: Probability of sub- and supercritical dynamics (red and blue, respectively; mean $\pm$ S.D.), obtained from applying the neural network to the independent holdout set. The crossing of both plots at $P=0.05$ indicates the critical point, matching the expected value of $T_c \approx 0.05$.}
\end{figure}

\begin{figure}
\includegraphics[scale=0.2]{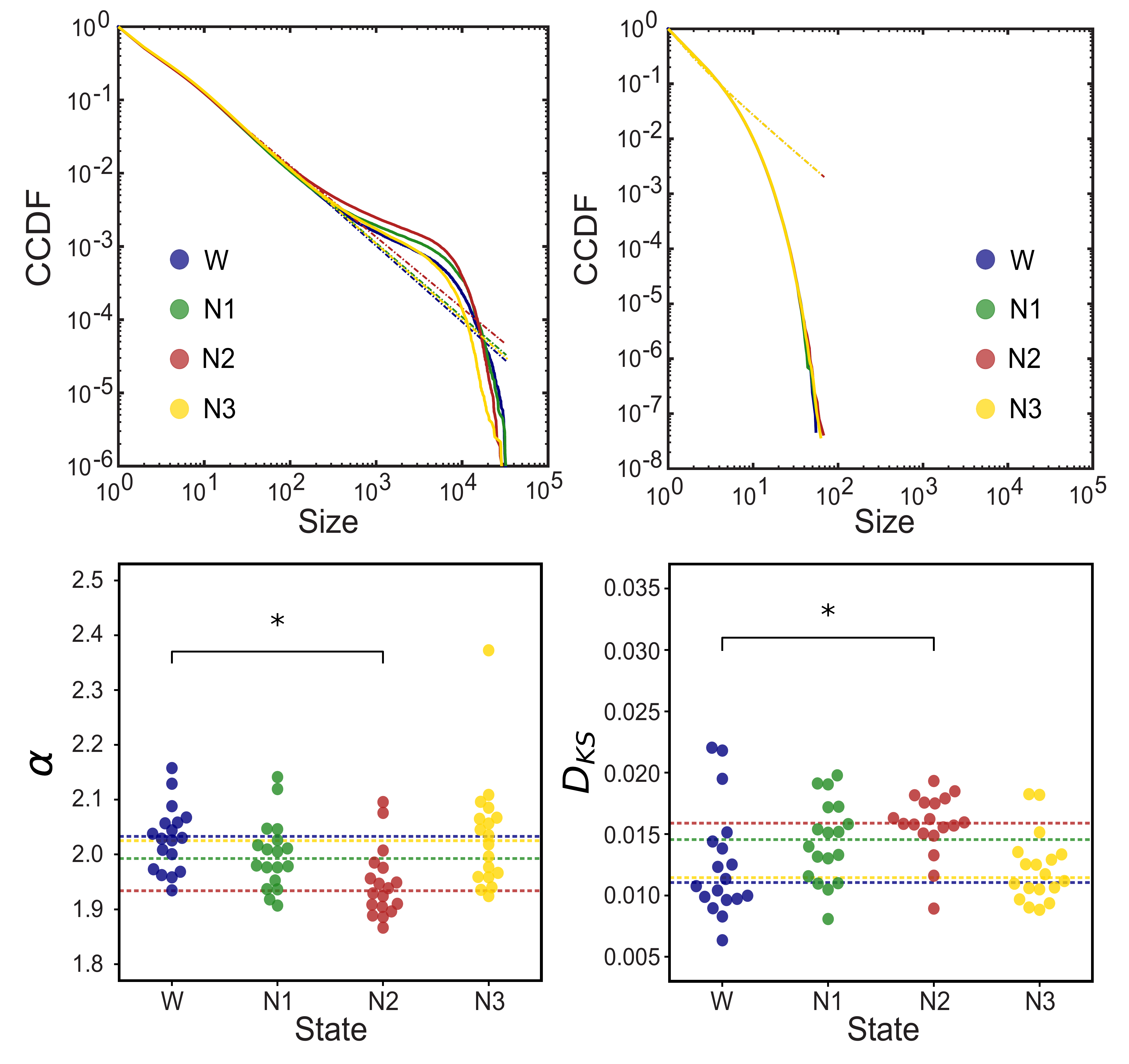}
\caption{\label{fmri} Scale-free dynamics in fMRI data. Upper left: complementary cumulative distribution functions (CCDF) for the size of connected clusters of active voxels obtained for wakefulness (W), and sleep (N1, N2, N3). Dotted lines correspond to power laws with exponents obtained using a maximum likelihood estimator applied to the empirical distributions. Upper right: same but obtained after randomly shuffling the phases of the voxel-wise signals. Lower left: swarm plots of the Fisher exponents $\alpha$ estimated for each participant and sleep stage (*p$<$0.05, Wilcoxon signed-rank test). Lower right: swarm plots of the goodness of fit, estimated as the Kolmogorov-Smirnov distance ($D_{KS}$) between the empirical distributions and the optimal fits (*p$<$0.05, Wilcoxon signed-rank test). All p-values were Bonferroni corrected for all pairwise comparisons between wakefulness and sleep states.}
\end{figure}

\emph{Brain activity model.—} We consider the Greenberg-Hastings model of excitable dynamics. The possible states and transition rules are shown in Fig. \ref{method}a. An inactive node $n_i$ can become active spontaneously with a small probability ($10^{-3}$), or if sum of the connection weights with its neighbouring active nodes is larger than a threshold, $\sum_{j} a_j W_{ij} > T$ where $a_j=1$ if $n_j$ is active and 0 otherwise. After becoming active, a node transition towards refractory in the next time step. For each time step, refractory node has a probability of $10^{-1}$ of returning to the inactive state. The model consists of 998 nodes connected according to the matrix $W_{ij}$ shown in Fig. \ref{method}b, obtained using DSI by Hagmann et al \cite{hagmann2008mapping}. The output of the model is a time series of binary (1 if active, 0 otherwise) vectors, representing the temporal evolution of the activations (Fig. \ref{method}c). 


\emph{Learning the phase transition.—} We implemented a fully connected deep neural network with an architecture illustrated in Fig. \ref{method}d. The input layer received the configuration of active nodes at each time step. This was followed by two hidden layers with 100 and 10 units, respectively, both with sigmoid activation functions, and by two softmax units in the output layer. The network was trained using L2 regularization with parameter equal to  0.01. To avoid overfitting, the connection between the hidden layers had a dropout rate of 0.5. 

This network was trained using examples simulated with the model using different $T$ values. Without spontaneous excitation, the Greenberg-Hastings model presents a phase transition at the point where activity becomes self-sustained \cite{fisch1993metastability}. As shown by Haimovici et al. and depicted in Fig. \ref{phase} (left), in the model with spontaneous excitation the phase transition can be described using $S_1$ (the size of the largest connected cluster) as order parameter, with $S_2$ peaking at the transition point (for this choice of $W_{ij}$ the critical value is $T_c \approx 0.05$) \cite{haimovici2013brain}.

To train the network, we simulated the model with $T$ ranging from 0.01 to 0.1 with step 0.001, starting from a configuration with 100 randomly active nodes and discarding the first 300 time steps to avoid transitory effects. For each $T$ value we simulated 1000 independent realizations, using 60\% of these to train the network and 40\% as a holdout set for evaluation. The network was trained to predict the phase, with $y_1$,$y_2$ the target value of the two output units and $y_1=1, y_2=0$ if  $T<T_c$, $y_1=0, y_2=1$ otherwise, using backpropagation with the Adam optimizer with 10 training epochs, and loss function given by the categorical cross-entropy, $-y_1 log(\hat{y}_1)-y_2 log(\hat{y}_2)$ ($\hat{y}_1, \hat{y}_2$ are the values of the two output units). As shown in Fig. \ref{phase} (right) the neural network is capable of predicting the critical point (crossing of both probabilities at $\approx$ 0.05) in the independent holdout set.


\emph{Power-law exponents in fMRI data.—} We thresholded each brain voxel in fMRI data with a threshold of +1 S.D. and obtained the distribution of sizes of activated clusters, $p(s)$ \cite{tagliazucchi2012criticality}. As shown in previous work \cite{tagliazucchi2012criticality, bocaccio2019avalanche}, this distribution follows a power law $p(s) \propto s^{-\alpha}$, with $\alpha \approx 2$ (Fisher exponent \cite{fisher1967theory}). We performed this analysis for fMRI data of 18 subjects during wakefulness (W) and all stages of the human sleep cycle (N1, N2 and N3, in increasing order of sleep depth). The complementary cumulative distribution functions (CCDF) for W, N1-N3 are shown in the left upper panel of Fig. \ref{fmri}. For the human fMRI data, power law scaling is apparent for cluster sizes smaller than $\approx 10^3$. The right upper panel of Fig. \ref{fmri} shows the CCDF obtained after randomly shuffling the phases of the voxel-wise fMRI time series, which effectively destroys the pair-wise correlations and results in a distribution indicative of exponential decay.

The scaling parameter $\alpha$ was estimated from the cumulative density function using a maximum-likelihood estimator which excluded values below a lower bound (estimated independently for each fit) \cite{clauset2009power}. As shown in the left lower panel of Fig. \ref{fmri}, the estimated $\alpha$ are in the range 1.8 - 2.1 (except for an outlier). There was a significant effect of stage (W, N1-N3) on $\alpha$ as determined using a Kruskal-Wallis test (p$<$0.05), and  $\alpha$ was significantly lower for N2 compared to W (p$<$0.05, Wilcoxon signed-rank test, Bonferroni corrected). The right lower panel shows the goodness of fit, obtained by computing the Kolmogorov-Smirnov distance ($D_{KS}$) between the empirical data and the best maximum likelihood fit. A Kruskal-Wallis test shown a significant effect between states (p$<$0.05), and the comparison between wakefulness (W) and sleep stages (N1-N3) determined significantly higher $D_{KS}$ values for N2 compared to W (p$<$0.05, Wilcoxon signed-rank test, Bonferroni corrected).

These changes suggested a progressive change in scale-free dynamics from wakefulness to N2 sleep, and a restoration during N3 sleep. The data was fitted equally well ($D_{KS} > 0.005$) using a log-normal, $P(s) \propto \frac{1}{s\sigma\sqrt{2\pi}}exp\left[-\frac{(ln (s)-\mu)^2}{2\sigma^2}\right]$, and a power law distribution with exponential cutoff, $P(s) \propto s^{-\alpha} e^{-\frac{s}{\beta}}$; however, the resulting parameters were consistent with an asymptotic limit where the distributions approach a power law ($\sigma>>1$ for the log-normal, in which case $\alpha \approx 1-\frac{\mu}{\sigma^2}$, and $\beta>>1$ for the power law with exponential cutoff). Finally, log-likelihood ratios were only marginally ($<3$) in favour of log-normal and power law with exponential cutoff, which are both models of higher complexity than the power law. 

Whether these results are indicative of a shift away from the critical point is more difficult to establish, given that scale-free dynamics is not an unambiguous signature of criticality \cite{girardi2021brain}. To further address this possibility, we turn to the comparison with the Greenberg-Hastings model using machine learning.

\begin{figure}
\includegraphics[scale=0.2]{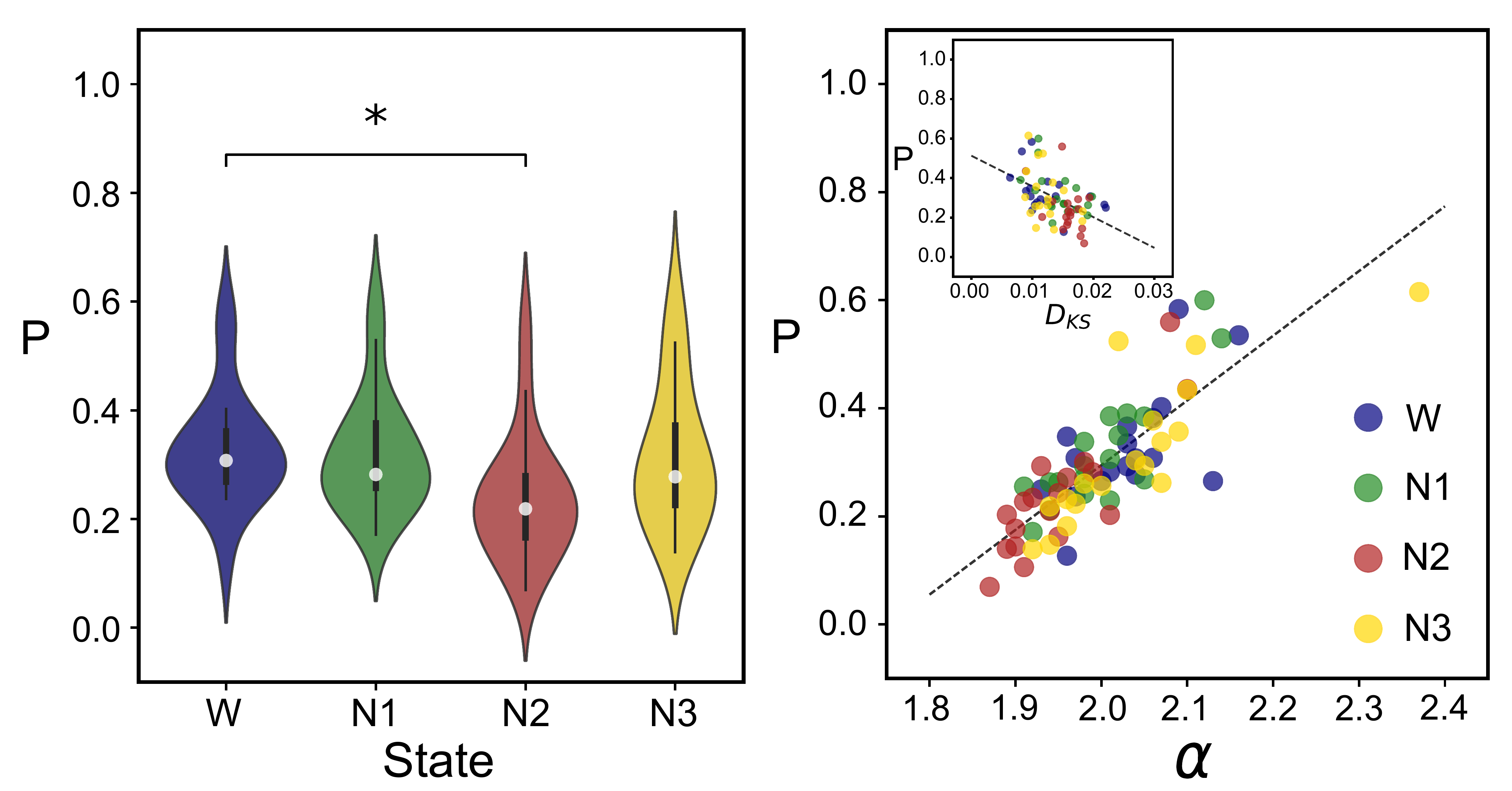}
\caption{\label{nnoutput} Model-based machine learning of critical brain dynamics. Left: Violin plots of the neural network output probabilities $P$ per sleep stage and participant (averaged across the $P$ obtained for all configurations corresponding to each sleep stage for each participant) (*p$<$0.05, Wilcoxon signed-rank test, Bonferroni corrected for all pairwise comparisons between wakefulness and sleep states). Right: Scatter plot of probabilities $P$ and power law exponents $\alpha$, one per subject and sleep stage. The dotted line indicates the best linear fit (least squares). A significant linear association between $P$ and $\alpha$ was found (R=0.81, p=1.3e-17). The inset shows a significant linear association between $P$ and $D_{ks}$ (R=-0.47, p=3.5e-5).}
\end{figure}

\emph{Correlation between exponents and predicted probabilities.—} We propose to use the neural network trained with the Greenberg-Hasting models to address this issue. We extracted the fMRI time series at the 998 locations defined by the parcellation in the DSI connectome of Fig. \ref{method}c, and binarized following the same criterion as before (threshold at +1 S.D.). This resulted in a sequence of binary vectors with 998 entries, which could be used as input to the neural network trained with similar data obtained simulating the model. Thus, for each time step of each subject and each sleep stage, the neural network provided an estimate of the probability of supercritical dynamics ($P$, with $P=0.5$ indicating proximity to the critical point). After averaging these values across all time points, a single $P$ value was obtained per subject and sleep stage. These values are summarized in the violin plots provided in Fig. \ref{nnoutput} (left). Again, we observed a significant effect of the stage (W, N1-N3) on $P$ (Krustal-Wallis test) and a significant different in $P$ between W and N2 sleep (Wilcoxon signed-rank test, Bonferroni corrected). Overall, the predicted $P$ values were indicative of subcritical dynamics, with N2 sleep being the most subcritical state. 

We conclude by showing that the changes in $\alpha$ (Fig. \ref{fmri}, lower left) could be explained by differences in the proximity to critical dynamics, according to the output of the neural network trained using the Greenberg-Hasting model. As shown in Fig. \ref{nnoutput} (right), the $\alpha$ values estimated as the power law exponents of the cluster size distributions were highly correlated with $P$, the probability that the configuration corresponds to the supercritical phase (R=0.81, p=1.3e-17). We also show in figure inset a significant but lower negative correlation (R=-0.47, p=3.5e-5) between P probabilities and Kolmogorov-Smirnov distance ($D_{KS}$). 


\emph{Discussion.—} Recent studies have shown that supervised and unsupervised machine learning algorithms are capable of learning phase transitions in a variety of physical systems \cite{wang2016discovering, van2017learning, carrasquilla2017machine, suchsland2018parameter}. Although phase transitions can be identified by a number of well understood signatures (typically associated with the appearance of scale-free distributions), none of these signatures are used directly as features to train the machine learning algorithms; instead, the information required to sort the phases of the system is inferred from the multivariate data. Here, we show that a neural network was not only capable of learning the phase transition in the Greenberg-Hastings model, but also reproduced changes in scaling parameters ($\alpha$) and goodness-of-fit statistics ($D_{KS}$) of power law fits when exposed to the full configuration vectors of the system. Consistently with previous reports \cite{priesemann2014spike, tomen2014marginally}, these changes are indicative of slightly subcritical dynamics, which are furthered in the progression from wakefulness to N2 sleep and restored during N3 sleep. This process could reflect the need to shift away from criticality until deep sleep is consolidated. 

Synthetic data is considered valuable in machine learning problems where the number of training samples is scarce and data augmentation is required \cite{perl2020data, van2001art}. We showed that simulated data alone can be used to train a model capable of predicting the proximity to the critical point in real empirical data, as suggested by the correlation with the changes in $\alpha$. This suggests a general strategy to identify the phases of physical systems based on empirical data, whereby simulations provide the information to represent these phases in the parameters of a deep neural network, which is then used to furnish predictions based on empirical data. Importantly, the model we explored is not based on a realistic formulation of neural activity (although it incorporates realistic connectivity), suggesting that universal aspects of dynamics near a phase transition suffice to train the network. It is also important to note that our results do not constitute proof of changes in the proximity to critical dynamics; however, the output of the neural network can be taken as an additional signature to assess this possibility, which is not directly based on features such as $\alpha$ or $D_{KS}$ but learned from the integral dynamics of the model.  

Specific applications of our development include the use of computational models to train algorithms capable of estimating levels of consciousness from brain imaging recordings, a problem of clinical relevance given the increasing number of patients who survive critical brain injury and enter a state of preserved vigilance, but with sporadic or inconsistent signs of conscious awareness \cite{gosseries2014recent}. For further validation we propose the application to physical systems theoretically known to undergo phase transitions that can be either controlled by the experimenter or inferred from multiple sources of high-quality empirical data. 
\vspace{10pt}

Authors acknowledge funding from Agencia I+D+i, Argentina (grant PICT-2019-02294) and ANID/FONDECYT Regular 1220995 (Chile).

\bibliographystyle{apsrev4-1} 

\bibliography{apssamp}

\end{document}


\preprint{PRL}

\title{Supplementary Material for ``Generative embeddings of brain collective dynamics using variational autoencoders''}


\author{Yonatan Sanz Perl}
  \affiliation{Universidad de San Andrés, Buenos Aires, Argentina}%
  \affiliation{Physics Department, University of Buenos Aires, and Buenos Aires Physics Institute, Argentina}%
   \affiliation{Center for Brain and Cognition, Computational Neuroscience Group, Universitat Pompeu Fabra, Spain}%
 \author{Hernán Bocaccio}%
\affiliation{Physics Department, University of Buenos Aires, and Buenos Aires Physics Institute, Argentina}%
\author{Ignacio Pérez-Ipiña}%
\affiliation{Physics Department, University of Buenos Aires, and Buenos Aires Physics Institute, Argentina}%
 \author{Federico Zamberlán}%
\affiliation{Physics Department, University of Buenos Aires, and Buenos Aires Physics Institute, Argentina}%
 \author{Juan Piccinini}%
\affiliation{Physics Department, University of Buenos Aires, and Buenos Aires Physics Institute, Argentina}%
  \author{Helmut Laufs}%
    \affiliation{Department of Neurology, Christian-Albrechts-University Kiel, Germany}
  \author{Morten Kringelbach}%
  \affiliation{Department of Psychiatry, University of Oxford, United Kingdom}%
  \author{Gustavo Deco}%
 \affiliation{Center for Brain and Cognition, Computational Neuroscience Group, Universitat Pompeu Fabra, Spain}%
 \author{Enzo Tagliazucchi}%
\affiliation{Physics Department, University of Buenos Aires, and Buenos Aires Physics Institute, Argentina}%

\date{\today}

\maketitle

\subsection{Sleep data set}
63 healthy subjects participated in the experiment (36 females, mean $\pm$ SD age of $23 \pm 43.3$ years). The experimental protocol was approved by the local ethics committee (Goethe-Universität Frankfurt, Germany, protocol number: $305/07$) and  written informed consent was asked to all participants before the experiments. All experiments were conducted in accordance with the relevant guidelines and regulations, including the Declaration of Helsinki.
The day of the study all participants reported a wake-up time between $5:00$ AM and $11:00$ AM, and a sleep onset time between $10:00$ PM and $2:00$ AM for the night prior to the experiment. Within half and hour of $7$ PM  participants entered the scanner and were asked to relax, close their eyes and not fight the onset of sleep. Their resting state activity was measured during $52$ minutes with a simultaneous combination of EEG and fMRI. 

With an optimized polysomnographic setting for sleep staging electroencephalography (EEG) and electromyography (EMG) were acquired.  According to the rules of the American Academy of Sleep Medicine \cite{berry2012aasm}, the scalp potentials measured with EEG determine the classification of sleep into $4$ stages (wakefulness, N1, N2 and N3 sleep). Previous publications based on this data set can be referenced for further details \cite{tagliazucchi2014decoding}.

EEG via a cap (modified BrainCapMR, Easycap, Herrsching, Germany) was recorded continuously during fMRI acquisition (1505 volumes of T2$^{*}$-weighted echo planar images, TR/TE = 2080 ms/30 ms, matrix 64$\times$64, voxel size 3$\times$3$\times$2 mm$^3$, distance factor 50\%; FOV 192 mm$^2$) with a 3 T Siemens Trio (Erlangen, Germany). An optimized polysomnographic setting was employed (chin and tibial EMG, ECG, EOG recorded bipolarly [sampling rate 5 kHz, low pass filter 1 kHz] with 30 EEG channels recorded with FCz as the reference [sampling rate 5 kHz, low pass filter 250 Hz]. Pulse oxymetry and respiration were recorded via sensors from the Trio [sampling rate 50 Hz]) and MR scanner compatible devices (BrainAmp MR+, BrainAmpExG; Brain Products, Gilching, Germany), facilitating sleep scoring during fMRI acquisition.

MRI and pulse artefact correction were performed based on the average artefact subtraction (AAS) method as implemented in Vision Analyzer2 (Brain Products, Germany) followed by objective (CBC parameters, Vision Analyzer) ICA-based rejection of residual artefact-laden components after AAS resulting in EEG with a sampling rate of 250 Hz. EEG artefacts due to motion were detected and eliminated using an ICA procedure implemented in Vision Analyzer 2.

\subsection{Anesthesia data set}
Resting state fMRI volumes from 18 healthy subjects were acquired in four different states following propofol injection: wakefulness, sedation, unconsciousness, and recovery. Data acquisition was performed in Liège (Belgium) (written informed consent, approval by the Ethics Committee of the Medical School of the University of Liège). Propofol was infused through an intravenous catheter placed into a vein of the right hand or forearm. An arterial catheter was placed into the left radial artery. Throughout the study, the subjects breathed spontaneously, and additional oxygen ($5$ $l/min$) was given through a loosely fitting plastic facemask. The level of consciousness was evaluated clinically throughout the study with the scale used in ref. \cite{maclaren2000prospective}. The subject was asked to strongly squeeze the hand of the investigator. She/he was considered fully awake or to have recovered consciousness if the response to verbal command (“squeeze my hand”) was clear and strong (Ramsay 2), as sedated if the response to verbal command was clear but slow (Ramsay 3), and as unconscious, if there was no response to verbal command (Ramsay 5–6). Ramsay scale verbal commands were repeated twice for each consciousness level assessment.  Functional MRI acquisition consisted of resting-state functional MRI volumes repeated in the four states: normal wakefulness (Ramsay 2), sedation (Ramsay 3), unconsciousness (Ramsay 5), and recovery of consciousness (Ramsay 2). The typical scan duration was half an hour for each condition, and the number of scans per session (200 functional volumes) was matched across subjects to obtain a similar number of scans in all states. Functional images were acquired on a 3 Tesla Siemens Allegra scanner (Siemens AG, Munich, Germany; Echo Planar Imaging sequence using $32$ slices; repetition time=2460 ms, echo time =40 ms, field of view =220 mm, voxel size $=3.45$\times$3.45$\times$3$ mm$^3$, and matrix size $=64$\times$64$\times$32$). 

\subsection{fMRI data preprocessing}
Using Statistical Parametric Mapping (SPM8, \url{www.fil.ion.ucl.ac.uk/spm}), Echo Planar Imaging (EPI) data were realigned, normalised (MNI space) and spatially smoothed (Gaussian kernel, 8 mm$^3$ full width at half maximum). Data was re-sampled to 4$\times$4$\times$4 mm resolution to facilitate removal of noise and motion regressors. Note that re-sampling introduces local averaging of Blood Oxygen Level Dependent (BOLD) signals, which were eventually averaged over larger cortical and sub-cortical regions of interest determined by the automatic anatomic labelling (AAL) atlas. Cardiac, respiratory (both estimated using the RETROICOR method and motion-induced noise (three rigid body rotations and translations, as well as their first 3 temporal derivatives, resulting in 24 motion regressors) were regressed out by retaining the residuals of the best linear least square fit. Data was band-pass filtered in the range 0.01–0.1 Hz using a sixth order Butterworth filter.

\subsection{Structural connectivity}

The structural connectome was obtained applying diffusion tensor imaging (DTI) to diffusion weighted imaging (DWI) recordings from 16 healthy right-handed participants (11 men and 5 women, mean age: 24.75 $\pm$ 2.54 years) recruited online at Aarhus University, Denmark. Subjects with psychiatric or neurological disorders (or a history thereof) were excluded from participation. The MRI data (structural MRI, DTI) were recorded in a single session on a 3T Skyra scanner. The following parameters were used for the structural MRI T1 scan: voxel size of 1 mm$^3$; reconstructed matrix size 256 $\times$ 256; echo time (TE) of 3.8 ms and repetition time (TR) of 2300 ms. DWI data were collected using the following parameters: TR = 9000 ms, TE = 84 ms, flip angle = 90, reconstructed matrix size of 106 $\times$ 106, voxel size of 1.98 $\times$ 1.98 mm with slice thickness of 2 mm and a bandwidth of 1745 Hz/Px. Furthermore, the data were recorded with 62 optimal nonlinear diffusion gradient directions at b = 1500 s/mm$^2$. Approximately one non-diffusion weighted image (b = 0) per 10 diffusion-weighted images was acquired. Additionally, the DTI images were recorded with different phase encoding directions. One set was collected applying anterior to posterior phase encoding direction and the second one was acquired in the opposite direction. The AAL template was used to parcellate the entire brain into 90 regions (76 cortical regions and 14 subcortical regions). The parcellation contained 45 regions in each hemisphere. To co-register the EPI image to the T1-weighted structural image, the linear registration tool from the FSL toolbox (\url{www.fmrib.ox.ac.uk/fsl}, FMRIB, Oxford) \cite{jenkinson2002improved} was employed. The T1-weighted images were co-registered to the T1 template of ICBM152 in MNI space. The resulting transformations were concatenated, inverted and further applied to warp the AAL template from MNI space to the EPI native space, where the discrete labeling values were preserved by applying nearest-neighbor interpolation. SC networks were constructed following a three-step process. First, the regions of the whole-brain network were defined using the AAL template. Second, the connections between nodes in the whole-brain network (i.e., edges) were estimated using probabilistic tractography for each participant (see below for more information). Third, results were averaged across participants.

Data preprocessing was performed using FSL diffusion toolbox (Fdt) with default parameters (\url{https://fsl.fmrib.ox.ac.uk/fsl/fslwiki/FDT}). Following this preprocessing, the local probability distributions of fiber directions were estimated at each voxel \cite{behrens2007probabilistic}. The probtrackx tool in Fdt was used to provide automatic estimation of crossing fibers within each voxel, which has been shown to significantly improve the tracking sensitivity of non-dominant fiber populations in the human brain \cite{behrens2007probabilistic}. After applying a Bayesian estimation of diffusion parameters (bedpostx Fdt tool), probtrackx2 starts from a seed voxel and produces streamlines by choosing an orientation (from the bedpostx distribution probability) and then takes a step in this direction, repeating until termination criteria are met. Repeating multiple times, the algorithm obtains a distribution of streamlines connecting a specific pair of regions, which represents the posterior distribution on the streamline location. Thus, the connectivity probability $P_{ij}$ from a seed voxel $i$ to another voxel $j$ was defined by the proportion of fibers passing through voxel $i$ that reached voxel $j$ (sampling of 5000 streamlines per voxel \cite{behrens2007probabilistic}). All the voxels in each AAL parcel were seeded (i.e. gray and white matter voxels were considered). This was extended from the voxel level to the region level, i.e. in a parcel consisting of $n$ voxels, $5000 \times n$ fibers were sampled. The connectivity probability  from region $i$ to region $j$ was calculated as the number of sampled fibers in region $i$ that connected the two regions, divided by  $5000 \times n$, where $n$ represents the number of voxels in region $i$. The resulting SC matrices were thresholded at 0.1\% (i.e. a minimum of five streamlines).

Due to the dependence of tractography on the seeding location, the probability from $i$ to $j$ was not necessarily equivalent to that from $j$ to $i$. However, these two probabilities were highly correlated across the brain for all participants (r $>$ 0.70, p $<$ 10$^{-50}$). As the directionality of connections cannot be determined using diffusion MRI, the unidirectional connectivity probability  between regions $i$ and $j$ was defined by averaging these two connectivity probabilities. This unidirectional connectivity was considered a measure of SC between the two areas, with $C_{ij} = C_{ji}$. The regional connectivity probability was calculated using in-house Perl scripts. For both phase encoding directions, 90 $\times$ 90 symmetric weighted networks were constructed based on the AAL parcellation, and normalized by the number of voxels in each AAL region, thus representing the SC network organization of the brain of each participant. Finally, the data was averaged across participants.

\subsection{Comparison of linear (PCA) vs. nonlinear (VAE) dimensionality reduction and justification for the use of VAE}

There are multiple reasons to represent the trajectory of brain states in a reduced latent space by means of a VAE, which at the same time exclude PCA as a viable alternative to achieve this purpose.

First, standard autoencoders are capable of achieving successful dimensionality reduction by means of nonlinear transformations of the input values (in our case, nonlinearities result from the use of rectified linear activation functions). Conversely, PCA introduces a new orthogonal system of coordinates and represents inputs as linear combinations of these coordinates; thus, in contrast to autoencoders, PCA can only be used for \emph{linear} dimensionality reduction. In other words, PCA can only learn a representation of the data in terms of \emph{hyperplanes}, while autoencoders are capable of learning an arbitrary nonlinear manifold. Since our models are inherently nonlinear we expected autoencoders to outperform PCA in terms of dimensionality reduction, which was confirmed in the subsequent analysis.

We applied PCA only using data from W and N3 sleep and identified the two-dimensional projection of the data along the two orthogonal directions associated with the largest eigenvalues. We then projected all samples from N1 and N2 sleep in this new coordinate system, obtaining the corresponding two-dimensional representation. This procedure is analogous to the one we followed to train the VAE, where the latent space representation of intermediate sleep stages emerged from training the network only with W and N3 data.
The left panel of Fig. S 1A shows the latent space representation of all samples obtained using a VAE, color coded to represent the four different sleep stages.  The right panel of Fig. S 1A shows the optimal results obtained from applying $1000$ iterations of the k-means algorithm to cluster the latent space representation into four clusters. The algorithm was capable of capturing the separation between W and N1, but confused part of the samples in the border between N3 and N2, yielding an average adjusted rand index of 0.88. Fig. S1B contains the results of the same analysis but using PCA instead of VAE. The left panel plots all samples in terms of their first two principal components. As opposed to the results obtained using VAE, the N1 and N2 clusters were not clearly separated from W and N3, respectively. Accordingly, the k-means algorithm failed to correctly group the samples belonging to W, N1, N2 and N3, yielding a lower adjusted rand index (0.75). 

We then  introduced a metric for the reconstruction error resulting from PCA and VAE. This metric consisted in computing the Euclidean distance between the lower triangular part of the original and reconstructed correlation matrices. To reconstruct input matrices using VAE we applied the decoder network to the latent space representation trained with samples from W and N3 sleep. To reconstruct the input matrices from the first two principal components we applied SVD to the samples $\times$ features matrix, and then produced a lower rank approximation by retaining only the two largest eigenvalues. Since this process included information from N1 and N2 sleep, it was biased towards a lower reconstruction error compared to VAE. However, in spite of this bias, the VAE reconstruction outperformed PCA for all sleep stages, as shown in Fig. S 1C. While the reconstruction error was comparable for W, PCA resulted in $\approx$ 30 $\%$  higher error for N3,  $\approx$ 150$\%$ for N1 and  $\approx$ 50 $\%$ for N2. 

\begin{figure*}[t]
\includegraphics[scale=0.75]{FigSupp_VAEvsPCA.pdf}
\caption{\emph{Comparison of linear (PCA) vs. nonlinear (VAE) dimensionality reduction}. A, left: All samples in the two-dimensional latent space determined by training a VAE with W and N3. Samples are color coded based on the sleep stage. A, right: Samples clustered using a k-means algorithm with four expected clusters, yielding an adjusted rand index of 0.88. B, left: Samples projected in the two-dimensional space spanned by the first two principal components obtained from W and N3 data. B, right: Samples clustered using a k-means algorithm with four expected clusters, yielding an adjusted rand index of 0.75. C, distributions for the reconstruction errors obtained from PCA and VAE. Errors were computed as the Euclidean distance between the lower triangular part of original and reconstructed correlation matrices.}
\end{figure*}

These results confirm that the linearity assumption is too stringent for the purpose of reducing the dimensionality of our data. Besides this limitation, there is another important reason that justifies the adoption of VAE instead of PCA. Units in the VAE bottleneck contain a latent space representation of the data and can be used to generate new samples which represent nonlinear combinations between the inputs used to train the network. This procedure was followed, for instance, in Fig. 2B and Fig. 2B of the manuscript, where we mapped the progression of sleep stages by exploring combinations of latent space units. Also, in Fig. 3A we showed that latent space units could be explored beyond this range to produce results with meaningful neurobiological interpretation (i.e. structure-function uncoupling when moving beyond N3 sleep).

Since standard PCA does not learn a probabilistic model of the data it is inadequate for this purpose. Probabilistic PCA assumes that the data is generated by a Gaussian latent variable model and then estimates the density of the latent variables \cite{tipping1999probabilistic}.  However, probabilistic PCA assumes a linear relationship between latent and observed variables, and precisely due to this limitation we chose VAE to learn the latent space representation of our data.  This decision is justified by the ubiquitous nonlinearities that characterize brain dynamics, and more directly by the results presented in Fig. 4 of our manuscript. Here, we show for a system of homogeneous Landau-Stuart oscillators that the latent space variables map approximately into linear combinations of the model parameters $a$ and $G$. However, the mapping between $a$ and $G$ and the output samples (i.e. correlation matrices) is guaranteed to be nonlinear due to the form of the dynamical equations (consider, for instance, the relationship between $a$ and the output correlation matrices as this parameter crosses the Hopf bifurcation).

While probabilistic PCA assumes a linear mapping from latent to sample space, there are other methods inspired by PCA which are capable of handling nonlinear mappings. Kernel PCA is a variant of probabilistic PCA where sample inputs are mapped into a new feature space by means of a nonlinear transformation, where standard probabilistic PCA can then be used to learn a nonlinear latent space manifold \cite{scholkopf1997kernel}. This nonlinear transformation can be complex and difficult to determine, but since the PCA algorithm only needs to compute the internal product between samples, then it is sufficient to determine the kernel of the transformation (a similar "trick” is used for SVM). VAE can be understood as a deep learning method to circumvent the problem of identifying the suitable kernel, with capacity for learning complex nonlinear kernels beyond frequent choices with a straightforward analytical expression (e.g. radial basis function or polynomial kernels). 

After comparing VAE with other linear/nonlinear and probabilistic/non-probabilistic methods, we conclude that VAE represents a suitable and adequate choice to learn the underlying structure of this kind of data.

\subsection{Latent space dimensionality justification}

We used a dynamical system to generate an ensemble of inputs for the VAE corresponding to different sleep stages. The underlying parameters of these models were optimized to maximize the similarity between simulated and empirical data, consisting of local perturbations ($\Delta_k$) to a homogeneous Hopf bifurcation parameter. Under the assumption that these perturbations represented comparatively small deviations from the homogeneous model, we decided for a two-dimensional latent space to capture the mean bifurcation parameter $a$ and the coupling scaling parameter $G$. Indeed, Fig. 4 shows how the VAE learns a combination of these two parameters when the $\Delta_k$ are all equal (homogeneous model).

Besides this justification in terms of the dynamical system used to generate our samples, we observe that these samples approximate instances of empirical fMRI connectivity matrices obtained during different sleep stages. Thus, there is also a neurobiological justification for our choice of latent space dimensionality. Both research scientists and clinicians frequently map the progression of sleep stages into a two-dimensional diagram consisting of vigilance/arousal (reactivity, wakefulness, capacity to react upon external stimulation) and awareness (reportable conscious mentation, in the case of sleep assessed off-line after awakening) \cite{laureys2009two}. While these two dimensions are usually correlated for human sleep stages (excluding REM sleep), deviations can appear, an example being illustrated in Fig. 2C, where N1 and W are set apart by the value of latent unit $z_1$ but not $z_2$. This is consistent with the speculation that unit $z_2$ roughly captures the “awareness” dimension (consider that very large negative values of neuron $z_2$ result in the kind of structural-functional uncoupling seen in deeply unconscious states such as general anesthesia) \cite{vincent2007intrinsic, barttfeld2015signature}. Also, this interpretation is consistent with the characterization of N1 as a state of reduced vigilance but preserved conscious awareness in the form of hypnagogic hallucinations, thought to be triggered by a delay between thalamic and cortical deactivation during sleep onset \cite{magnin2010thalamic}. While we are not claiming that units $z_1$ and $z_2$ neatly map into these two dimensions, we nevertheless support their joint relevance to provide a neurobiologically meaningful characterization of the data, thus justifying our choice of latent space dimensionality.

\subsection{Comparison between empirical and simulated correlation matrices}

Figure S \ref{comp} presents a comparison between the average empirical ($C_{ij}^{emp}$, above diagonal) and simulated ($C_{ij}$, below diagonal) correlation matrices. The SSIM similarity index is comparable to that reported in previous works, as well as to the values obtained from the Greenberg-Hastings model (see Fig. S4). The main difference between matrices $C_{ij}^{emp}$ and $C_{ij}$ is the presence of strong homotopic (i.e. interhemispheric) correlations in the empirical data, manifest as high values along the contradiagonal of the matrix. This discrepancy is known to arise due to the underestimation of long-rage fiber tracts by DTI probabilistic tractography \cite{reveley2015superficial}, in particular, the long tracts that connect pairs of homotopic regions tend to be underestimated \cite{messe2014relating}. While ad hoc methods could be applied to increase the similarity (such as artificially adding a fixed positive value to the matrix entries that correspond to homotopic pairs) \cite{ipina2020modeling} we opted to work with unaltered DTI matrices. Since our results were obtained in spite of this difference between empirical and simulated matrices, we expect that revisiting our analysis when better DTI estimates become available will strengthen our conclusions.

\begin{figure*}[t]
\includegraphics[scale=0.35]{FigSupp_Hopf.pdf}
\caption{\label{comp} \emph{Comparison between empirical and simulated correlation matrices}. The four matrices present a comparison between the average empirical ($C_{ij}^{emp}$, above diagonal) and simulated ($C_{ij}$, below diagonal) correlation matrices among different stages of sleep cycle. The SSIM similarity index is comparable to that reported in previous works, as well as to the values obtained from the Greenberg-Hastings model (see Fig. S4). The main difference between matrices $C_{ij}$ and $C_{ij}^{emp}$  is the presence of strong homotopic (i.e. interhemispheric) correlations in the empirical data, manifest as high values along the contradiagonal of the matrix. This discrepancy is known to arise due to the underestimation of long-rage fiber tracts by DTI probabilistic tractography \cite{reveley2015superficial}, in particular, the long tracts that connect pairs of homotopic regions tend to be undestimated. While ad hoc methods could be applied to increase the similarity (such as artificially adding a fixed positive value to the matrix entries that correspond to homotopic pairs) \cite{ipina2020modeling} we opted to work with unaltered DTI matrices. Since our results were obtained in spite of this difference between empirical and simulated matrices, we expect that revisiting our analysis when better DTI estimates become available will strengthen our conclusions.}
\end{figure*}

\subsection{Two-dimensional representation of sleep stages by means of the optimised model parameters}

We have investigated whether the sleep stages presented separations or clear trajectories in terms of the model parameters representing the contribution of each RSN to the local bifurcation parameter ($\Delta_k$). Fig. S \ref{deltas} presents two-dimensional scatter plots between all pairs of parameters $\Delta_k$. Each point corresponds to optimal parameter obtained after fitting the Stuart-Landau model to the empirical correlation matrices, color-coded for the corresponding sleep stage. The lack of a clear separation or trajectory between states is evident upon direct visual inspection (compare with the results in latent space, Fig. 2A of the main manuscript text). We quantified this using the k-means clustering algorithm with four expected clusters and then computing the adjusted Rand index between the cluster labels and the sleep stage classification (1000 iterations, insets show mean $\pm$ STD). All Rand index values were significantly smaller than the result obtained when applying the same procedure to the latent space representation in Fig. 2A (0.88).

\begin{figure*}[t]
\includegraphics[scale=0.65]{FigSupp_parameters.pdf}
\caption{\label{deltas} \emph{Two-dimensional representation of sleep stages by means of the optimised model parameters}. We have investigated whether the sleep stages presented separations or clear trajectories in terms of the model parameters that representing the contribution of each RSN to the local bifurcation parameter ($\Delta_k$). Panels in the figure presents two-dimensional scatter plots between all pairs of parameters $\Delta_k$. Each point corresponds to optimal parameter obtained after fitting the Stuart-Landau model to the empirical correlation matrices, color coded for the corresponding sleep stage. The lack of a clear separation or trajectory between states is evident upon direct visual inspection (compare with the results in latent space, Fig. 2A of the main manuscript text). We quantified this using the k-means clustering algorithm with four expected clusters and then computing the adjusted Rand index between the cluster labels and the sleep stage classification ($1000$ iterations, insets show mean $\pm$ SD). All Rand index values were significantly smaller than the result obtained when applying the same procedure to the latent space representation in Fig. 2A (0.88). Vis (visual), Aud (auditory), SM (sensorimotor), DMN (default mode network), EC (executive control).}

\end{figure*}

\subsection{Biological interpretation of the latent space organization}

N1 sleep is a transitional (i.e. not stable) stage between wakefulness and N2 sleep. Of all non REM sleep stages, the prevalence of N1 sleep during a complete night of sleep is the lowest (5\% compared to 50\% in N2 and 20\% in N3) \cite{carskadon2011normal}. Furthermore, since N1 sleep is mainly defined by a drop in EEG alpha (8-12 Hz) power and an increase in theta (4-8 Hz), it can be very challenging to detect in individuals who have low baseline alpha amplitude \cite{berry2012aasm}. An experienced sleep scorer might score 30 s. of EEG (as per the standards of the American Academy of Sleep Medicine) \cite{berry2012aasm} as wakefulness, and then suddenly detect a K-complex in the signal, i.e. a large amplitude ($>$100 $\mu$V) peak in the EEG which indicates the start of N2 sleep. In this case, the apparent absence of N1 sleep would lead the scorer to revise the previous 30 s. epochs and introduce a cutoff in alpha power to avoid a sudden transition between W and N2, especially if the recording corresponds to the early part of the night (see below). In some cases, the jump between W and N2 sleep is so pronounced that N1 sleep does not appear in the hypnogram. Also, human REM sleep is divided in cycles lasting $\approx$90 minutes each. After the first cycle, it is common to “skip” N1 sleep and to transition directly into N2 sleep. This is supports that the latent space organization was capable of inferring a relevant feature of human sleep architecture.

\subsection{Replication using a different model for data augmentation}

We speculate that the model is capturing a property of the Stuart-Landau model which is simultaneously a generic property of other dynamical systems and a signature of brain dynamics as they transition from wakefulness into deep sleep or other states characterized by reduced awareness and vigilance: a transition between ordered (e.g. oscillatory, synchronous) and disordered (e.g. noisy, uncorrelated) dynamics \cite{steyn2005sleep}. Fitting local bifurcation parameters to the Stuart-Landau model to reproduce empirical correlation matrices evidenced this transition \cite{ipina2020modeling}. In the following, we investigate this hypothesis by reproducing the embedding with a different model that presents similar behavior.
We implemented a discrete time model presenting an order-disorder transition (Greenberg-Hastings model) \cite{haimovici2013brain}. In this model, each node in the DTI network is modeled as a unit with three possible states: inactive, active and refractory. The inactive state can turn into an active state either spontaneously with a small probability (10$^{-3}$), or if enough first neighbors are activated. In the next step after activation, the node becomes refractory (i.e. immune to activation) and for each time step has a certain probability of returning to the baseline state (10$^{-1}$). This model represents a macroscopic version of the inactive-depolarized-hyperpolarized cell cycle, and includes an order-disorder transition with the order parameter given by the activity lifetime. The parameters and equations describing the temporal evolution of the model are provided in Fig. S \ref{GH}A, and the procedure followed to obtain the empirical correlation matrices is explained in the caption. Figure S \ref{GH}B presents the comparison of empirical (above diagonal, $C_{ij}^{emp}$) and simulated (below diagonal, $C_{ij}$) correlation matrices. As with the Stuart-Landau model, the main difference appears in the contra diagonal, which reflects the underestimation of homotopic connectivity by DTI \cite{messe2014relating, reveley2015superficial, ipina2020modeling}. As the Stuart-Landau model, the Greenberg-Hastings model captures the gradual loss of correlations from W to N3, with the simultaneous preservation of certain correlations between nodes that are strongly connected in the anatomical sense.

The results of encoding N1 and N2 correlation matrices using a VAE trained with samples from W and N3 is shown in Fig. S \ref{homo} (left). After training the neural network, correlation matrices simulated from the Greenberg-Hastings model fitted to the empirical data were encoded into the latent units $z_1$ and $z_2$. This figure shows that (as in the case of the Stuart-Landau model) samples from W and N3 are clearly separated. Samples from N1 and N2 tend to be closer to W than for the Stuart-Landau model, but nevertheless show discernible centroids, with N2 sleep presenting lower values of $z_2$ and higher values of $z_1$, and N1 sleep presenting higher values of $z_2$. Thus, using the Greenberg-Hastings model do not map into the exact same manifold as those obtained using the Stuart-Landau model, but the VAE was capable of isolating the intermediate states, and embedding their centroids into a low dimensionality manifold. While the precise configuration of this embedding might depend on the chosen model, the VAE nevertheless was capable of mapping the sequence of brain states into an orderly progression within this particular manifold.
The right panel of Fig. S \ref{homo} presents the correlation matrices obtained by exploring the latent space, showing a gradient that resembles Fig. 2B (Stuart-Landau model).
Thus, we have shown that another model with comparable qualitative behavior (order-disorder phase transition) maps similarly into the latent space of the VAE. This supports that our results are not the consequence of specific features of the Stuart-Landau model. In line with several previous reports applying phenomenological models to reproduce large-scale brain activity patterns \cite{cofre2020whole}, even if these two models are not biologically realistic, their dynamics and second order statistics nevertheless reproduce fMRI observables computed from the stages of human sleep, and thus can be used as data augmentation devices to train the VAE. 

\begin{figure*}[t]
\includegraphics[scale=0.25]{FigSupp_New2ModelsV1.pdf}
\caption{\label{GH} \emph{Whole-brain model with local dynamics based on Greenberg-Hastings model}. A. Illustration of the spreading mechanism in the Greenberg-Hasting model. Each node in the DTI network can be in one of three states (inactive, active and refractory). Transitions from inactive to active occur either spontaneously with a small probability ($10^{-3}$), or if the sum of the connection weights $W_{ij}$ from the given node to active neighbouring nodes exceeds a threshold T. After activation, nodes transition into a refractory state which is immune to activation. Finally, nodes recover to the inactive state with a fixed probability ($10^{-1}$). Simulated correlation matrices were computed after time series binarization and convolution with an hemodynamic response function kernel (\cite{haimovici2013brain}). These matrices were fitted to the empirical data following the same procedure applied to the Stuard-Landau model, in particular, we allowed for regional variability in the threshold parameter T (spatially constrained by the RSN).  B. Comparison of empirical (above diagonal, $C_{ij}^{emp}$) and simulated (below diagonal, $C_{ij}$ ) correlation matrices. As with the Stuart-Landau model, the main difference appears in the contradiagonal, which reflects the underestimation of homotopic connectivity by DTI (\cite{messe2014relating,reveley2015superficial,ipina2020modeling})}
\end{figure*}

\begin{figure*}[t]
\includegraphics[scale=0.25]{FigSupp_CA_Multi_latent.pdf}
\caption{\label{homo} \emph{Latent space sleep stage organisation obtained with Greenberg-Hastings model}. Left: Latent state representation obtained by encoding the test set (W and N3) and the encoding obtained for the two intermediate states that were not used to train the VAE (N1 and N2 sleep), with correlation matrices generated using the Greenberg-Hastings model fitted to the empirical data. Right: Correlation matrices obtained by decoding an exhaustive exploration of the latent space variables $z1$ and $z2$. }
\end{figure*}

\subsection{Replication using a different sequence of brain states}

 We used the same model (Stuart-Landau) to generate samples from different brain states, and to show that the autoencoder trained using W and N3 data produced an embedding consistent with the known neurophysiology of those states. We followed the same procedure to fit the model to data acquired under propofol sedation and during propofol-induced loss of consciousness (for details on the experimental procedures followed for fMRI acquisition under propofol, see \cite{tagliazucchi2016large}). These states are characterized by changes in brain dynamics consistent with those observed during intermediate and deep sleep \cite{murphy2011propofol}. Furthermore, the neurochemical pathways activated by propofol anesthesia present overlap with those responsible for initiating and maintaining sleep \cite{nelson2002sedative}. We then expected these states to be encoded in the proximity of N3 sleep, which is confirmed in Fig. S \ref{propo}.

\begin{figure*}[t]
\includegraphics[scale=0.5]{FigSupp_prop2.pdf}
\caption{\label{propo}  \emph{Propofol sedation and propofol-induced loss of consciousness representation in latent space}. We followed the same procedure to fit the model to data acquired under propofol sedation and during propofol-induced loss of consciousness (for details on the experimental procedures followed for fMRI acquisition under propofol, please see \cite{schrouff2011brain} ). These states are characterized by changes in brain dynamics consistent with those observed during intermediate and deep sleep \cite{murphy2011propofol}. Furthermore, the neurochemical pathways activated by propofol anesthesia present overlap with those responsible for initiating and maintaining sleep \cite{nelson2002sedative}. The figure shows the encoding of correlation matrices corresponding to propofol sedation (\emph{Sed}) and loss of consciousness (\emph{Inc}) in a latent space obtained training a VAE with samples from W and N3 sleep. As we expected these states are encoded in the proximity of N3 sleep. }
\end{figure*}

\subsection{Reproduction of the main results using the lower triangular part of the correlation matrices}

The 8100 matrix entries contain redundant information. We expected that this redundant information would not affect the results, since the properly trained autoencoder would learn to reproduce the full matrices using only information from the upper/lower triangular part. We show that this is indeed the case in Fig. S \ref{lower}, which reproduces Fig. 2 of the manuscript using only the inferior triangular part of the correlation matrices (note that some variability is expected due to the stochastic nature of the variational autoencoder).

\begin{figure*}[t]
\includegraphics[scale=0.5]{Fig2_rev1.pdf}
\caption{\label{lower}  \emph{Reproduction of the main results using the lower triangular part of the correlation matrices}. Considering that the full matrices present redundant information, we repeated the procedure to obtain Fig. 2 of the manuscript using only the lower triangular part. We expected that this redundant information would not affect the results, since the properly trained autoencoder would learn to reproduce the full matrices using only information from the upper/lower triangular part. We show in this figure that this is indeed the case, reproducing Fig. 2 of the manuscript using only the inferior triangular part of the correlation matrices (note that some variability is expected due to the stochastic nature of the variational autoencoder). }
\end{figure*}

\subsection{Mean and variance of the correlation matrices generated in the latent space}

The left panel of Fig. S \ref{meanvar} presents the mean of the correlation matrix throughout the latent space, while the right panel presents the variance. There is a gradient in the mean spanned by a linear combination of $z_1$ and $z_2$. The relationship between the variance and the latent space units is more complex, with a minimum for a particular combination of values.

\begin{figure*}[t]
\includegraphics[scale=0.4]{var_mean.png}
\caption{\label{meanvar}  \emph{Mean and variance of the correlation matrices generated in the latent space.}  The left panel presents the mean of the correlation matrix throughout the latent space, while the right panel presents the variance. There is a gradient in the mean spanned by a linear combination of $z_1$ and $z_2$. The relationship between the variance and the latent space units is more complex, with a minimum for a particular combination of values.}
\end{figure*}

\clearpage
\bibliographystyle{apsrev4-1} 

\bibliography{apssamp}